\RequirePackage{fix-cm}
\documentclass[twocolumn,epjc3]{svjour3}  
\usepackage{amsmath}
\usepackage{amssymb}
\smartqed  
\RequirePackage{graphicx}
\journalname{Eur. Phys. J. C}
\begin{document}

\title{Quasinormal Modes of Self-Dual Loop Quantum Gravity Corrected Kerr Black Holes
}

\author{Yu Wang\thanksref{e1,addr1,addr2}\and Meilin Liu\thanksref{e2,addr2}
\and Haiguang Xu\thanksref{addr1}}
\thankstext{e1}{e-mail: sjtu2686361@sjtu.edu.cn (Yu Wang)}
\thankstext{e2}{Corresponding author: meilin.liu@sjtu.edu.cn (Meilin Liu)}

\institute{School of Physics and Astronomy, Shanghai Jiao Tong University, Shanghai 200240, China \label{addr1}
           \and
           School of Aeronautics and Astronautics, Shanghai Jiao Tong University, Shanghai 200240, China \label{addr2}
}
\date{\today}
\maketitle

\begin{abstract}
We investigate scalar quasinormal modes and ringdown signals of a rotating loop quantum gravity black hole constructed via the revised Newman--Janis algorithm. The spacetime incorporates two quantum parameters, the polymeric parameter $P$ and the minimal area parameter $a_0$, and reduces to the Kerr spacetime in the classical limit. Using the sixth-order WKB approximation and time-domain integration, we analyze the effective potential, quasinormal mode spectrum, and ringdown dynamics. The polymeric parameter $P$ systematically reduces the oscillation frequency $\mathrm{Re}(\omega)$, while the spin parameter $a$ enhances it. The time-domain signals exhibit clear dependence on the multipole number $\ell$ and overtone index $n$. Our results provide a theoretical framework for understanding the effects of loop quantum gravity corrections on black hole perturbations and gravitational-wave ringdown signals.
\end{abstract}


\section{Introduction}

Quasinormal modes (QNMs) describe the characteristic oscillations of black holes under external perturbations~\cite{Kokkotas1999,Nollert1999,Berti2009,Konoplya2011}. 
They correspond to a discrete set of complex frequencies determined solely by the background spacetime and the nature of the perturbing field. 
The real part of the frequency characterizes the oscillation frequency, while the imaginary part determines the damping rate of the perturbation.
In addition, deviations from the classical QNM spectrum can serve as probes of exotic compact objects and new physics beyond general relativity~\cite{Cardoso2016}.

Since the first direct detection of gravitational waves from binary black hole mergers by LIGO and Virgo~\cite{LIGO2016,LIGO2019}, the ringdown phase has become a powerful observational window into strong-field gravity. 
During this stage, the gravitational-wave signal is dominated by QNMs, which encode the mass, spin, and geometry of the remnant black hole~\cite{Kerr1963,Teukolsky1973}. 
This provides a unique opportunity to test the no-hair property of general relativity and to search for possible deviations from the Kerr paradigm.

The QNM spectra of classical black holes, particularly Schwarzschild and Kerr solutions, have been extensively studied both analytically and numerically~\cite{ReggeWheeler1957,Chandrasekhar1983,Leaver1985}. 
Despite their success, general relativity is expected to break down near the Planck scale, where quantum gravitational effects become relevant. 
Such effects may modify the effective spacetime geometry and, consequently, the perturbation potential governing QNMs, leading to observable deviations in the ringdown spectrum.

Among candidate theories of quantum gravity, loop quantum gravity (LQG) provides a nonperturbative and background-independent quantization of spacetime~\cite{Rovelli2004,Thiemann2007,Ashtekar2004}. 
In this framework, geometric quantities such as area and volume acquire discrete spectra, introducing a fundamental quantum length scale. 
Effective black hole spacetimes inspired by LQG have been explicitly constructed in several regular models that capture leading quantum corrections while preserving classical consistency~\cite{Modesto2004,Modesto2006,Modesto2010,Gambini2013,Bodendorfer2013}. 
The canonical quantization of spherically symmetric LQG black holes has been systematically developed, establishing the fundamental operator structures and constraint analysis~\cite{Thiemann1993,Campiglia2007,Bengtsson1988}. 
Key geometric aspects, including the Hamiltonian constraint, volume operator, and quantum horizon structures, have been rigorously studied within the spherically symmetric reduction~\cite{Bojowald2000,Bojowald2004,Bojowald2005,Bojowald2006,Kuchar1994}.

In particular, self-dual black hole solutions introduce quantum corrections through polymeric parameters and minimal area effects, resulting in regular geometries free from curvature singularities~\cite{Sakharov2005,Barrau2014,Barrau2016}. 
These models provide a phenomenological framework for exploring potential observational signatures of quantum gravity.

Most existing studies focus on static and spherically symmetric LQG black holes. 
However, astrophysical black holes are expected to possess non-negligible angular momentum. 
Therefore, it is important to construct rotating extensions of LQG-inspired black holes. 
A powerful approach is the Newman--Janis algorithm, which generates rotating solutions from static spacetimes while preserving asymptotic properties~\cite{NewmanJanis1965,Azreg2014}.
Recently, several rotating LQG black hole solutions have been constructed and their QNM spectra have been investigated~\cite{LQGShadow2020,LQGQNM2022,LQGSpin2023,EPJCLQG2024,Zhu2025PRD,RotatingLQG2026}.

In recent years, increasing attention has been devoted to testing quantum gravity effects through gravitational-wave observations. 
Future detectors such as LISA, the Einstein Telescope, and Cosmic Explorer are expected to significantly improve the precision of ringdown measurements~\cite{LISA2017,EinsteinTelescope,CosmicExplorer}. 
This opens the possibility of constraining deviations from Kerr geometry and probing quantum corrections through QNM spectra.

In this work, we investigate scalar quasinormal modes of a rotating loop quantum gravity black hole constructed via the Newman--Janis algorithm. 
Starting from a static self-dual LQG black hole, we derive the rotating spacetime and compute the effective potential for scalar perturbations. 
The quasinormal frequencies are obtained using a higher-order WKB approximation. 
We analyze how the LQG polymeric parameter and the black hole spin modify the effective potential and shift the QNM spectrum, providing potential observational signatures of quantum gravitational effects.

The paper is organized as follows. In Section 2, we review the rotating self-dual black-hole spacetime and discuss the influence of the quantum parameters on the horizon structure. In Section 3, we derive the scalar perturbation equation and reduce it to a Schr"odinger-like wave equation with an effective potential. In Section 4, we present the WKB approximation method used to compute the quasinormal frequencies and briefly discuss its validity in the present context. In Section 5, we provide a detailed numerical analysis of the quasinormal mode spectrum and investigate the effects of the spin parameter and quantum corrections. In Section 6, we perform time-domain simulations and analyze the corresponding ringdown signals. Finally, Section 7 summarizes our main results and discusses possible future extensions, including gravitational perturbations and observational constraints from gravitational-wave detectors.

\section{Loop Quantum Gravity black holes and the revised Newman--Janis construction}

Loop quantum gravity (LQG) is a nonperturbative and background-independent approach to quantum gravity, in which geometric operators such as area and volume possess discrete spectra. This leads to the emergence of a minimal length scale at the Planck regime and provides a natural mechanism to regularize classical curvature singularities. In the semiclassical limit, effective spacetime descriptions can be constructed that incorporate leading-order quantum gravitational corrections while preserving the classical manifold structure.

A representative example is the self-dual loop quantum black hole (LQBH), which provides a quantum-corrected extension of the Schwarzschild spacetime. The static LQBH metric is written as
\begin{equation}
ds^2 = -G(r)\,dt^2 + F^{-1}(r)\,dr^2 + H(r)\,d\Omega^2,
\end{equation}
with
\begin{equation}
d\Omega^2 = d\theta^2 + \sin^2\theta\, d\phi^2.
\end{equation}

The metric functions are given by
\begin{equation}
G(r) =
\frac{(r-r_+)(r-r_-)(r+r_*)^2}{r^4 + a_0^2},
\end{equation}

\begin{equation}
F(r) =
\frac{(r-r_+)(r-r_-)r^4}{(r+r_*)^2 (r^4 + a_0^2)},
\end{equation}

\begin{equation}
H(r) = r^2 + \frac{a_0^2}{r^2},
\end{equation}
where
\begin{equation}
r_+ = 2M,\qquad r_- = 2M P^2,\qquad r_* = 2M P.
\end{equation}

Here $P$ is the polymeric parameter characterizing quantum geometric corrections, while $a_0$ represents the minimal area scale. In the classical limit $P \to 0$ and $a_0 \to 0$, the Schwarzschild spacetime is recovered.

To construct the rotating extension, we employ the revised Newman--Janis algorithm (NJA). Unlike the original complexification method, the revised NJA generates an axisymmetric spacetime directly from the static seed metric while preserving asymptotic flatness and avoiding ambiguities in mass identification.

The resulting rotating loop quantum black hole metric takes the form
\begin{equation}
\begin{aligned}
ds^2 = {} &
-\left(1 - \frac{2m r}{\rho^2}\right) dt^2
- \frac{4 a m r \sin^2\theta}{\rho^2} dt\, d\phi \\
& + \frac{\rho^2}{\Delta} dr^2
+ \rho^2 d\theta^2
+ \frac{\Sigma \sin^2\theta}{\rho^2} d\phi^2,
\end{aligned}
\end{equation}

where the metric functions are
\begin{equation}
\rho^2 = K(r) + a^2 \cos^2\theta,
\end{equation}

\begin{equation}
\Delta = F(r)\, r^2 + a^2,
\end{equation}

\begin{equation}
\Sigma = \left(K(r) + a^2\right)^2 - a^2 \Delta \sin^2\theta,
\end{equation}

and the deformation function is
\begin{equation}
K(r) = r^2 \sqrt{\frac{F(r)}{G(r)}}.
\end{equation}

Here $a = J/M$ is the spin parameter, and $m$ is the ADM mass of the spacetime. In the asymptotic region $r \to \infty$, the metric satisfies
\begin{equation}
M_{\rm ADM}
= \lim_{r \to \infty} \frac{r}{2}\left(F(r)^{-1} - 1\right)
= m,
\end{equation}
ensuring consistency with the Schwarzschild limit.

In the limit $P \to 0$ and $a_0 \to 0$, the Kerr metric is recovered, while in the non-rotating limit $a \to 0$, the static LQBH solution is obtained. This rotating loop quantum black hole provides a consistent framework to investigate quantum gravitational corrections in rotating spacetimes. In particular, the parameters $P$ and $a_0$ modify the near-horizon geometry and are expected to modify the effective potential, quasinormal modes, and ringdown properties.

\section{Scalar Perturbation Equation and Effective Potential}

We consider a massless scalar field $\Phi$ propagating in the spacetime of a rotating loop quantum black hole constructed via the revised Newman--Janis algorithm. The dynamics of the scalar field is governed by the covariant Klein--Gordon equation,
\begin{equation}
\nabla_\mu \nabla^\mu \Phi = 0,
\end{equation}
where $\nabla_\mu$ denotes the covariant derivative associated with the rotating LQBH metric.

To perform the mode decomposition, we adopt the ansatz
\begin{equation}
\Phi(t,r,\theta,\phi)
=
e^{-i\omega t}
Y_{\ell m}(\theta,\phi)
\frac{\psi(r)}{r},
\end{equation}
where $Y_{\ell m}(\theta,\phi)$ are the spherical harmonics and $\psi(r)$ is the radial function describing scalar perturbations. In the slow-rotation regime, the spin-zero spheroidal harmonics can be approximated by the spherical harmonics at leading order.

In the present work, we adopt the slow-rotation approximation $a\omega \ll 1$. Under this approximation, the coupling between the rotation parameter $a$ and the azimuthal quantum number $m$ appears at higher order $\mathcal{O}(am\omega)$ and is neglected at leading order. Therefore, the perturbation equation retains the radial structure of the corresponding spherically symmetric case, and the dominant contribution depends mainly on the multipole number $\ell$.

Under this approximation, the radial perturbation equation can be rewritten in a Schr\"odinger-like form,
\begin{equation}
\frac{d^2 \psi}{dr_*^2}
+
\left[\omega^2-V_\ell(r)\right]\psi=0,
\end{equation}
where the tortoise coordinate $r_*$ is defined by
\begin{equation}
dr_*=\frac{r^2+a^2}{\Delta(r)}\,dr,
\end{equation}
with $\Delta(r)$ determined by the rotating loop quantum black hole geometry obtained through the revised Newman--Janis algorithm.

The effective potential governing scalar perturbations is given by
\begin{equation}
V_\ell(r)
=
\frac{\Delta(r)}{(r^2+a^2)^2}
\left[
\ell(\ell+1)
+\frac{2M(r)r}{r^2+a^2}
\right],
\end{equation}
where $M(r)$ denotes the effective mass function induced by the loop quantum gravity corrections through the revised Newman--Janis construction. The function $\Delta(r)$ encodes both the rotational contribution and the quantum corrections inherited from the underlying LQBH geometry.

Although this effective potential is derived within the slow-rotation approximation, it captures the essential features of scalar perturbations in the rotating loop quantum black hole spacetime. A fully consistent treatment of rapidly rotating perturbations would require the Teukolsky formalism, which is beyond the scope of the present work.

The quasinormal mode spectrum is mainly determined by the behavior of the effective potential near its maximum. Denoting the peak position by $r_0$, the maximum value of the potential is given by
\begin{equation}
V_0=V_\ell(r_0).
\end{equation}
The corresponding quasinormal frequencies are calculated using the WKB approximation, which depends on both the height and curvature of the effective potential at the peak.

In particular, the damping behavior of the quasinormal modes is mainly governed by the second derivative of the potential at the peak,
\begin{equation}
\left.
\frac{d^2V_\ell}{dr_*^2}
\right|_{r=r_0}.
\end{equation}

We further investigate the effects of the black hole spin parameter $a$, the polymeric parameter $P$, and the minimal area parameter $a_0$ on the effective potential. The spin parameter mainly controls the frame-dragging effect and shifts the peak of the potential toward smaller radii. In contrast, the quantum parameters $P$ and $a_0$ modify the radial structure of the metric functions inherited from the LQBH geometry, leading to deviations from the classical Kerr potential and changing the height and width of the effective potential barrier.

The multipole number $\ell$ determines the strength of the centrifugal barrier. Larger values of $\ell$ increase the height of the effective potential and lead to a more localized scattering region.

Overall, both the rotational effects and loop quantum gravity corrections modify the effective potential structure. Since the quasinormal frequencies are mainly determined by the properties of the potential near its maximum, these modifications can induce characteristic changes in the gravitational-wave ringdown signals.

\section{Quasinormal Modes of the Rotating LQG-Corrected Black Hole}

In this section, we investigate the quasinormal modes (QNMs) of the rotating loop quantum gravity (LQG)-corrected black hole spacetime. Quasinormal modes describe the characteristic damped oscillations of perturbed black holes and play an important role in black hole spectroscopy. The corresponding complex frequencies are determined by the background geometry and therefore provide a useful tool for studying quantum gravitational corrections in the strong-field regime.

We consider scalar perturbations propagating in the rotating LQG-corrected spacetime introduced in Sec.~2. For simplicity, we focus on massless scalar perturbations governed by the Klein--Gordon equation,
\begin{equation}
\Box \Psi=0 .
\end{equation}

Using the mode decomposition
\begin{equation}
\Psi(t,r,\theta,\phi)
=
e^{-i\omega t}
Y_{\ell m}(\theta,\phi)
\frac{\psi(r)}{r},
\end{equation}
where $Y_{\ell m}$ denotes the spherical harmonics and $\omega$ is the quasinormal frequency, the perturbation equation can be reduced, under the slow-rotation approximation, to a Schr\"odinger-like radial wave equation,
\begin{equation}
\frac{d^2\psi}{dr_*^2}
+
\left[\omega^2-V_\ell(r)\right]\psi=0 .
\label{waveeq}
\end{equation}

The tortoise coordinate is defined as
\begin{equation}
dr_*=\frac{r^2+a^2}{\Delta(r)}dr ,
\end{equation}
where $\Delta(r)$ is determined by the rotating LQG black hole geometry obtained through the revised Newman--Janis algorithm.

The effective potential for scalar perturbations is given by
\begin{equation}
V_\ell(r)
=
\frac{\Delta(r)}
{(r^2+a^2)^2}
\left[
\ell(\ell+1)
+
\frac{2M(r)r}{r^2+a^2}
\right],
\label{potential}
\end{equation}
where $M(r)$ represents the effective mass function induced by the loop quantum gravity corrections. The function $\Delta(r)$ contains both the rotational contribution and the quantum modifications of the underlying LQG geometry.

The quasinormal modes satisfy the boundary conditions of purely ingoing waves at the event horizon and purely outgoing waves at spatial infinity,
\begin{equation}
\psi(r_*)\sim e^{-i\omega r_*},
\qquad r_*\rightarrow-\infty ,
\end{equation}
and
\begin{equation}
\psi(r_*)\sim e^{+i\omega r_*},
\qquad r_*\rightarrow+\infty .
\end{equation}

To calculate the quasinormal frequencies, we employ the sixth-order Wentzel--Kramers--Brillouin (WKB) approximation. The WKB expansion gives
\begin{equation}
i\frac{\omega^2-V_0}
{\sqrt{-2V_0''}}
-\sum_{j=2}^{6}\Lambda_j
=
n+\frac{1}{2},
\label{wkb}
\end{equation}
where $V_0$ is the maximum value of the effective potential, $V_0''$ denotes the second derivative of the potential with respect to the tortoise coordinate evaluated at the peak, $n$ is the overtone number, and $\Lambda_j$ represent higher-order WKB correction terms.

The QNM spectrum is therefore determined by the properties of the effective potential near its maximum. The variations of the spin parameter $a$ and the quantum correction parameters $P$ and $a_0$ modify the height and curvature of the potential barrier, leading to corresponding changes in the real and imaginary parts of the quasinormal frequencies.

The real part of the frequency $\omega_R$ represents the oscillation frequency of the ringdown signal, while the imaginary part $\omega_I$ determines the damping rate,
\begin{equation}
\omega=\omega_R-i\omega_I .
\end{equation}

The loop quantum gravity corrections modify the radial structure of the rotating black hole geometry through the functions $\Delta(r)$ and $M(r)$, and consequently change the effective potential in Eq.~(\ref{potential}). In particular, the height, width, and peak position of the potential barrier depend on the quantum parameters $P$ and $a_0$, as well as the spin parameter $a$. These modifications lead to characteristic shifts in both the oscillation frequencies and damping rates of the quasinormal modes.

To investigate the effects of the LQG corrections, we numerically calculate the quasinormal frequencies for different values of the parameters $P$, $a_0$, and $a$. Our results show that the quantum parameters deform the effective potential barrier and induce systematic variations in the QNM spectrum. In particular, increasing the polymeric parameter $P$ generally reduces the oscillation frequency by modifying the potential structure, while the minimal area parameter $a_0$ produces comparatively milder corrections. The spin parameter $a$ mainly affects the potential peak through rotational effects and increases the oscillatory behavior of the quasinormal modes.

The imaginary part of the frequency is also modified by the combined effects of rotation and quantum corrections, indicating changes in the damping behavior of the ringdown signal. These deviations become more significant in the strong-field region, where the quantum corrections to the spacetime geometry are enhanced.

An important aspect of black hole spectroscopy is the connection between quasinormal modes and unstable null geodesics. In the eikonal limit $\ell\gg1$, the quasinormal frequencies are approximately related to the properties of unstable circular photon orbits through
\begin{equation}
\omega_{\rm QNM}
\approx
\ell\,\Omega_c
-i\left(n+\frac{1}{2}\right)\lambda ,
\end{equation}
where $\Omega_c$ is the angular frequency of the unstable circular null orbit and $\lambda$ is the corresponding Lyapunov exponent. Since the LQG corrections modify the photon orbit structure, they naturally induce deviations from the classical Kerr quasinormal spectrum.

\begin{table}[htbp]
\centering
\caption{Quasinormal modes of the rotating LQG black hole calculated using the sixth-order WKB approximation.}
\label{tab:lqg_qnm}
\begin{tabular}{ccccc}
\hline
$\ell$ & $n$ & $r_0$ & $\mathrm{Re}(\omega)$ & $\mathrm{Im}(\omega)$ \\
\hline
2 & 0 & 2.7987 & 0.522060 & -0.033811 \\
2 & 1 & 2.7987 & 0.514270 & -0.102970 \\
2 & 2 & 2.7987 & 0.499967 & -0.176526 \\
3 & 0 & 2.8180 & 0.719412 & -0.033067 \\
3 & 1 & 2.8180 & 0.713931 & -0.099963 \\
3 & 2 & 2.8180 & 0.703413 & -0.169097 \\
\hline
\end{tabular}
\end{table}

The ringdown waveform can be approximately expressed as
\begin{equation}
h(t)\sim e^{-\omega_I t}
\cos(\omega_R t+\phi_0),
\end{equation}
where $\phi_0$ is a constant phase. Therefore, the modifications of $\omega_R$ and $\omega_I$ induced by the spin and quantum parameters lead to characteristic changes in the oscillation frequency and decay behavior of the ringdown signal.

These results demonstrate that the analysis of quasinormal mode spectra provides an effective approach to probing loop quantum gravity corrections in rotating black hole spacetimes. Future gravitational-wave detectors with improved sensitivity may constrain the quantum parameters through black hole spectroscopy and precise measurements of ringdown signals.

\section{Effective potential and parameter dependence}

We study the effective perturbation potential in a LQG--Kerr spacetime characterized by the rotation parameter $a$ and the quantum deformation parameter $P$. The effective potential takes the form
\begin{equation}
V(r)=\frac{\Delta(r)}{(r^2+a^2)^2}
\left[l(l+1)+\frac{2Mr}{r^2+a^2}\right], \quad l=2,
\end{equation}
where the radial function is modified by a LQG-inspired deformation,
\begin{equation}
\Delta(r)=r^2 F_{\mathrm{LQG}}(r;P)+a^2.
\end{equation}

This construction preserves the Kerr angular structure while introducing quantum-gravity motivated corrections in the radial sector.

The effective potential exhibits a single-barrier structure, with the peak corresponding to an unstable photon-sphere-like orbit. To characterize its properties, we extract three key observables: the peak height $V_{\max}$, the peak location $r_{\max}$, and the full width at half maximum (FWHM).

\begin{figure}[t]
\centering
\includegraphics[width=0.48\textwidth]{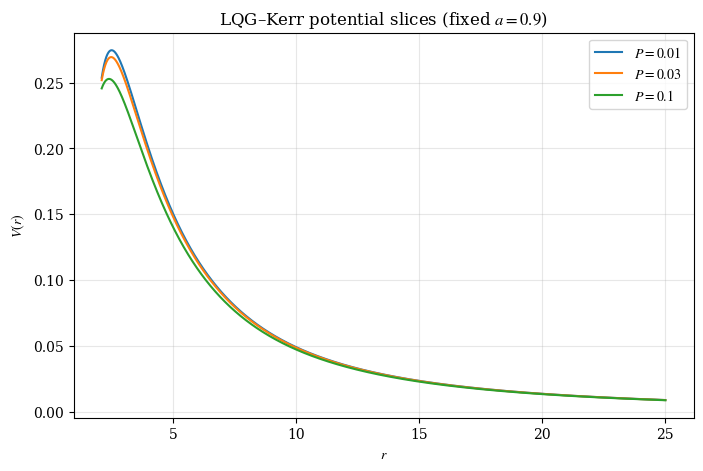}
\caption{
Radial profiles of the LQG--Kerr effective potential for fixed spin $a=0.9$ and different values of the deformation parameter $P$.
Increasing $P$ smooths and broadens the potential barrier while slightly reducing its peak amplitude.
}
\label{fig:radial}
\end{figure}

To first understand the local structure of the effective potential, we show representative radial profiles for fixed spin $a=0.9$ and varying $P$ in Fig.~\ref{fig:radial}.

From Fig.~\ref{fig:radial}, one observes that increasing $P$ systematically smooths the potential barrier and increases its width, while the peak amplitude decreases moderately. This behavior is consistent with a quantum-gravity motivated regularization of the near-horizon structure.

To further quantify the interplay between $a$ and $P$, we construct two-dimensional parameter maps of $V_{\max}$, $r_{\max}$, and FWHM in the $(a,P)$ plane.

\begin{figure*}[t]
\centering
\includegraphics[width=0.95\textwidth]{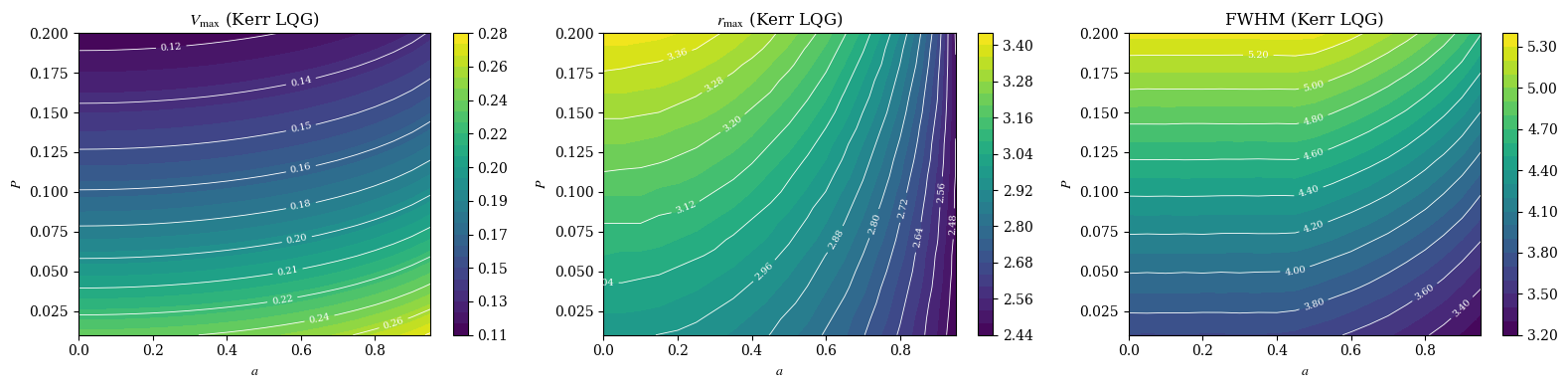}
\caption{
Two-dimensional parameter dependence of the LQG--Kerr effective potential characteristics.
Left: peak height $V_{\max}$.
Middle: peak location $r_{\max}$.
Right: FWHM.
The spin parameter $a$ primarily controls the geometric deformation of spacetime, while the LQG parameter $P$ governs the radial smoothing of the potential barrier.
}
\label{fig:ap_contour}
\end{figure*}

The numerical results shown in Fig.~\ref{fig:ap_contour} reveal a clear separation of the effects of the spin parameter $a$ and the LQG deformation parameter $P$ on the effective potential.

As $P$ increases, the peak height $V_{\max}$ systematically decreases, while the peak location $r_{\max}$ remains almost unchanged. At the same time, the potential barrier becomes noticeably broader, as reflected by an increase in the FWHM. This indicates that the LQG parameter $P$ primarily acts to weaken and smooth the effective potential barrier without significantly shifting its radial position.

In contrast, the spin parameter $a$ produces an opposite trend in the peak height: $V_{\max}$ increases with increasing $a$. Meanwhile, the peak position $r_{\max}$ shifts slightly toward smaller radii, indicating an inward displacement of the effective potential barrier due to stronger frame-dragging effects. The barrier width also exhibits a mild increase with $a$, suggesting a weak broadening of the scattering region in the rotating case.

Overall, the results demonstrate that $P$ mainly controls the amplitude suppression and radial smoothing of the effective potential, whereas $a$ governs the enhancement of the peak height and a slight inward shift of the barrier location. These distinct behaviors provide a partial separation between quantum-gravity-induced deformations and classical rotational effects in the Kerr LQG spacetime.

\begin{figure*}[t]
\centering
\includegraphics[width=0.95\linewidth]{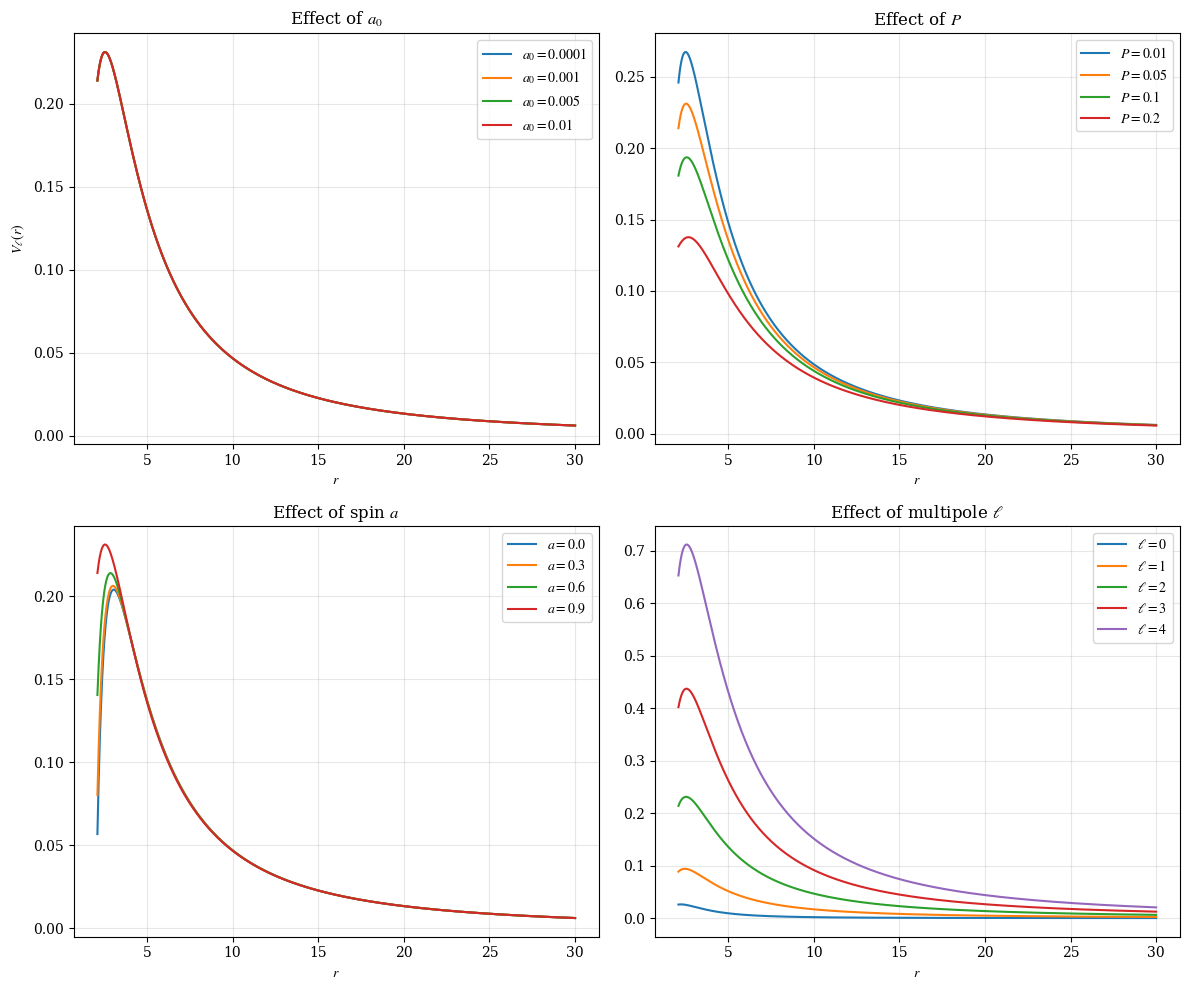}
\caption{
Effective potential $V_\ell(r)$ in the NJLQG spacetime.
Upper-left: dependence on $a_0$.
Upper-right: dependence on $P$.
Lower-left: dependence on spin $a$.
Lower-right: dependence on multipole number $\ell$.
}
\label{fig:potential_structure}
\end{figure*}

To systematically investigate the structure of the effective potential in the NJLQG spacetime, we analyze its dependence on the quantum parameter $a_0$, the deformation parameter $P$, the spin $a$, and the multipole number $\ell$. The results are summarized in Fig.~\ref{fig:potential_structure}.

In the upper-left panel, we observe that increasing $a_0$ leads to a slight suppression of the potential height, while the overall shape remains largely unchanged. This indicates that $a_0$ introduces only a mild quantum correction at the level of the effective potential.

In the upper-right panel, the effect of $P$ is more pronounced: increasing $P$ significantly lowers the peak of the potential and broadens the barrier, indicating a clear smoothing effect of loop quantum gravity corrections on the strong-field structure.

The lower-left panel shows that the spin parameter $a$ enhances the potential barrier height and slightly shifts the peak position toward smaller radii, reflecting stronger frame-dragging effects in the rotating background.

Finally, the lower-right panel demonstrates the expected multipole behavior: higher $\ell$ modes produce higher and sharper potential barriers, consistent with the centrifugal contribution in the effective potential.

\subsection{Parameter Dependence of Kerr--LQG Quasinormal Modes}

In this subsection, we systematically investigate the dependence of the quasinormal mode (QNM) frequencies on the black hole spin parameter $a$, the LQG polymeric parameter $P$, and the multipole number $\ell$. The overtone number is fixed to $n=0$.

\begin{figure*}[htbp]
\centering
\includegraphics[width=0.95\textwidth]{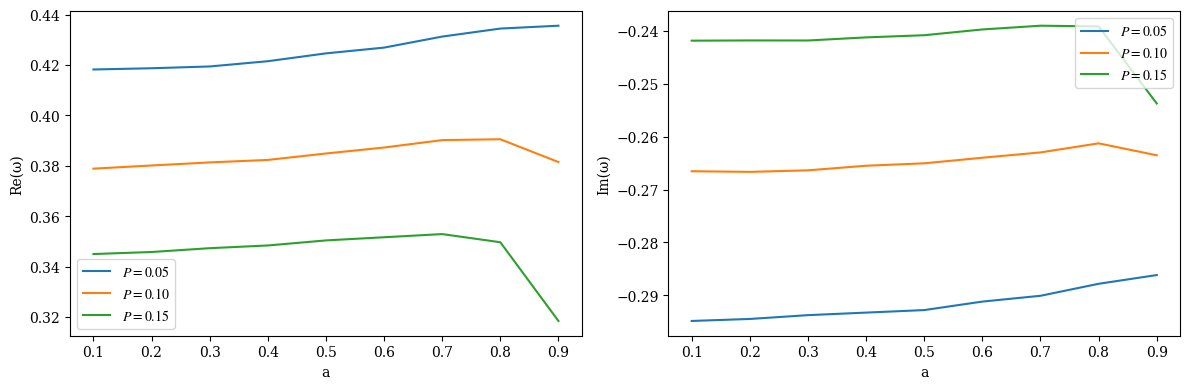}
\caption{
Dependence of Kerr--LQG quasinormal mode frequencies on the rotation parameter $a$.
Left panel: real part $\mathrm{Re}(\omega)$; right panel: imaginary part $\mathrm{Im}(\omega)$.
Different curves correspond to different values of the LQG parameter $P$.
The multipole number is fixed to $\ell=2$ and the overtone number to $n=0$.
}
\label{fig:qnm_a}
\end{figure*}

\begin{figure*}[htbp]
\centering
\includegraphics[width=0.95\textwidth]{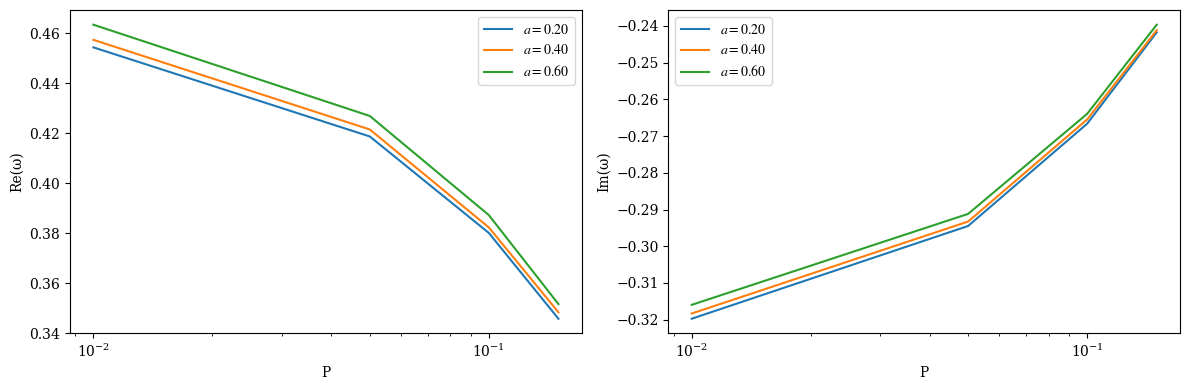}
\caption{
Dependence of Kerr--LQG quasinormal mode frequencies on the LQG polymeric parameter $P$.
Left panel shows $\mathrm{Re}(\omega)$, while the right panel shows $\mathrm{Im}(\omega)$.
Different curves correspond to different fixed values of the spin parameter $a$.
A logarithmic scale is used for the $P$ axis to highlight the weak-coupling regime.
}
\label{fig:qnm_P}
\end{figure*}

\begin{figure*}[htbp]
\centering
\includegraphics[width=0.95\textwidth]{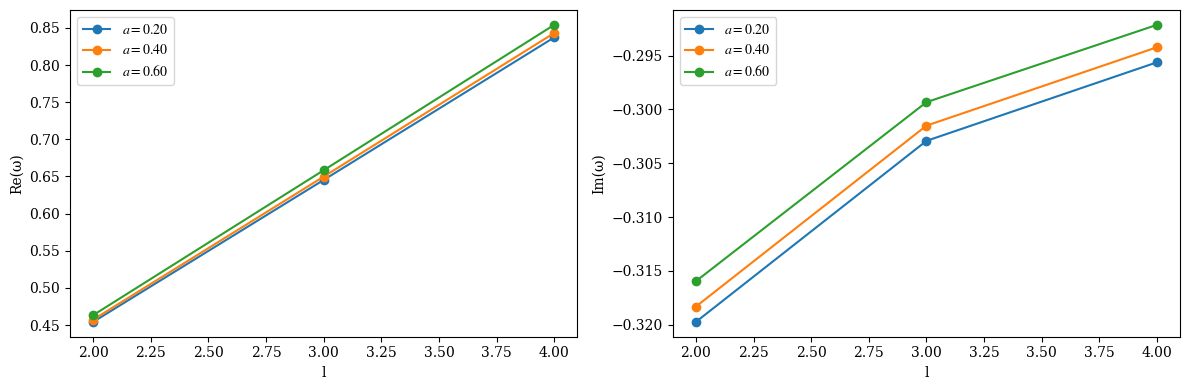}
\caption{
Dependence of Kerr--LQG quasinormal mode frequencies on the multipole number $\ell$.
Left panel: $\mathrm{Re}(\omega)$; right panel: $\mathrm{Im}(\omega)$.
Different curves correspond to different values of the spin parameter $a$.
The LQG parameter is fixed to $P=0.1$ and the overtone number is $n=0$.
}
\label{fig:qnm_l}
\end{figure*}

Figure~\ref{fig:qnm_a}, Figure~\ref{fig:qnm_P}, and Figure~\ref{fig:qnm_l} present the behavior of the real and imaginary parts of the QNM frequencies under variations of $a$, $P$, and $\ell$, respectively.

In Fig.~\ref{fig:qnm_a}, we observe that the real part of the frequency $\mathrm{Re}(\omega)$ generally increases with the spin parameter $a$, indicating that rotation enhances the characteristic oscillation scale of the perturbations. The imaginary part $\mathrm{Im}(\omega)$ shows a more subtle behavior: for small and moderate values of $a$, the damping rate decreases slightly with increasing spin, corresponding to a mild suppression of the decay. 

However, at sufficiently large spin, this trend begins to reverse, and the damping rate starts to increase. This turnover behavior becomes more pronounced for larger values of the LQG parameter $P$, suggesting a nontrivial interplay between quantum gravitational corrections and rotation in the strong-field regime.

In Fig.~\ref{fig:qnm_P}, we show the dependence of the QNM spectrum on the LQG correction parameter $P$. We find that both $\mathrm{Re}(\omega)$ and $\mathrm{Im}(\omega)$ decrease smoothly as $P$ increases. In particular, the oscillation frequency exhibits a stronger sensitivity to $P$, indicating that quantum gravitational corrections primarily reduce the effective oscillation scale of the perturbations. The damping rate also decreases with increasing $P$, suggesting a slower decay of the quasinormal modes in the presence of stronger LQG effects.

In Fig.~\ref{fig:qnm_l}, we present the dependence of the QNM frequencies on the multipole number $\ell$. We find that higher multipole modes lead to an increase in the oscillation frequency, while producing a slight decrease in the damping rate. This trend remains robust across different values of the spin parameter $a$, indicating that it is a generic feature of the Kerr--LQG perturbation spectrum.

In addition, we observe that the oscillation frequency increases with the spin parameter $a$, whereas the damping rate decreases as the black hole rotates faster. This behavior suggests that rotation enhances the oscillation scale of the perturbations while slightly suppressing the decay rate of the quasinormal modes.

Overall, these results indicate that the Kerr--LQG QNM spectrum exhibits a clear and systematic dependence on both classical (rotation, multipole number) and quantum (polymeric parameter) effects, making it a promising probe for potential quantum gravitational signatures in black hole ringdown observations.

\subsection{Parameter Dependence of Quasinormal Modes in Kerr--LQG Spacetime}

In this subsection, we analyze the dependence of the quasinormal mode (QNM) spectrum on the loop quantum gravity (LQG) correction parameter $P$ and the black hole rotation parameter $a$. The multipole and overtone numbers are chosen as $\ell=2,3$ and $n=0,1,2$.

\begin{figure*}[htbp]
\centering
\includegraphics[width=0.95\textwidth]{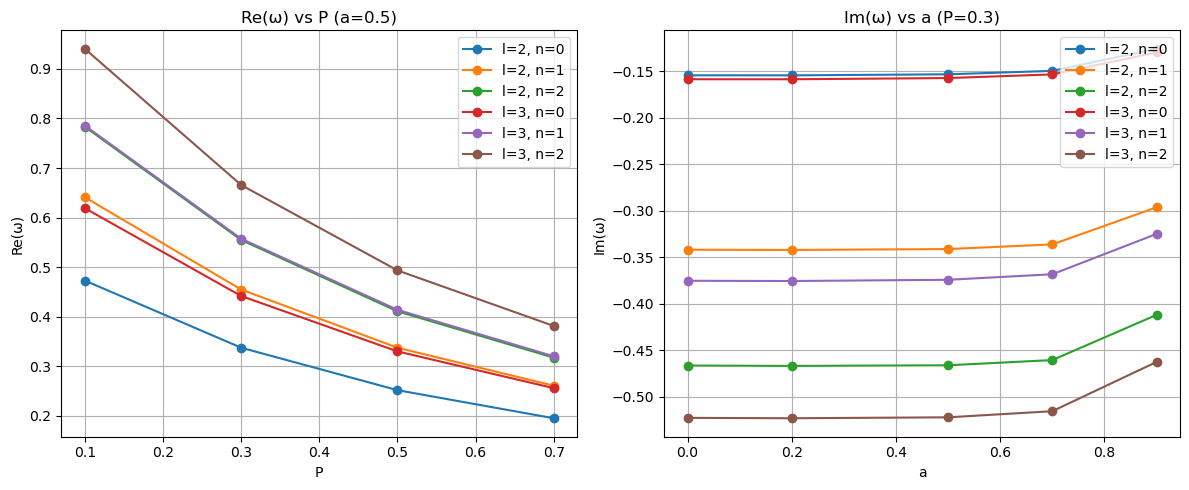}
\caption{
Dependence of quasinormal mode frequencies in the Kerr--LQG spacetime.
Left panel: real part of the frequency $\mathrm{Re}(\omega)$ as a function of the LQG parameter $P$ for fixed $a=0.5$.
Right panel: imaginary part $\mathrm{Im}(\omega)$ as a function of the spin parameter $a$ for fixed $P=0.3$.
Different curves correspond to different combinations of multipole and overtone numbers $(\ell, n)$.
}
\label{fig:lqg_qnm_scan}
\end{figure*}

Figure~\ref{fig:lqg_qnm_scan} shows the variation of the real and imaginary parts of the QNM frequencies as functions of the LQG parameter $P$ and the spin parameter $a$, respectively. In the left panel, the real part of the frequency $\mathrm{Re}(\omega)$ is plotted as a function of $P$ for fixed spin $a=0.5$. In the right panel, the imaginary part $\mathrm{Im}(\omega)$ is shown as a function of $a$ for fixed $P=0.3$.

We observe that the oscillation frequency decreases systematically as the LQG parameter $P$ increases, indicating that quantum gravitational corrections reduce the characteristic oscillation scale of the perturbations. In contrast, the damping rate exhibits only a weak dependence on the spin parameter $a$. Over a broad range of spins, the imaginary part of the quasinormal frequency remains nearly unchanged, while a slight reduction in the damping rate is observed for rapidly rotating black holes.

This behavior is consistently found for different multipole and overtone numbers, suggesting that the observed trends are generic features of perturbations in the Kerr--LQG spacetime rather than artifacts associated with specific modes. Although higher overtones remain more sensitive to the near-horizon geometry, the dominant influence of the polymeric parameter is reflected in the reduction of the oscillation frequency, whereas the effect of rotation on the damping rate remains comparatively modest.

Overall, these results indicate that the polymeric parameter $P$ primarily modifies the oscillation frequency of the quasinormal modes, while the black-hole spin has only a limited impact on the damping behavior, producing a slightly weaker damping for sufficiently large values of $a$.
\subsection{Quasinormal Mode Trajectories in the Complex Plane for Kerr--LQG Black Holes}

In this subsection, we investigate the impact of loop quantum gravity (LQG) corrections on the quasinormal mode (QNM) spectrum of rotating black holes by analyzing the trajectories of the complex frequencies in the $\omega$-plane. The black hole multipole number, overtone number, and mass are fixed to $\ell=2$, $n=0$, and $M=1$, respectively.

\begin{figure}[htbp]
\centering
\includegraphics[width=0.45\textwidth]{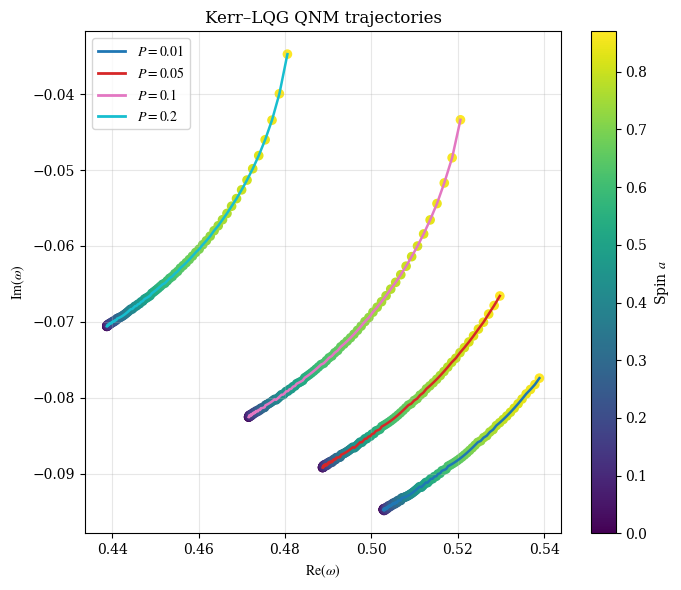}
\caption{
Complex plane trajectories of quasinormal mode frequencies for Kerr--LQG black holes.
Each curve corresponds to a fixed value of the polymeric parameter $P$,
while the rotation parameter $a$ is continuously varied from $0$ to $0.99$.
The color gradient indicates the variation of $a$ along each trajectory.
The results are shown for $P = 0.01, 0.05, 0.1, 0.2$ with fixed parameters $\ell=2$, $n=0$, and $M=1$.
}
\label{fig:LQG_QNM_trajectory}
\end{figure}

Figure~\ref{fig:LQG_QNM_trajectory} shows the evolution of the QNM frequencies for different values of the polymeric parameter $P = 0.1, 0.3, 0.5, 0.7$. For each fixed value of $P$, the rotation parameter $a$ is continuously varied from $0$ to $0.99$, generating a parametric trajectory of the complex frequency $\omega = \omega_R + i \omega_I$ in the complex plane.

We observe that both the real part $\omega_R$ and the imaginary part $\omega_I$ exhibit a systematic deformation of their trajectories as the LQG parameter $P$ increases. In particular, larger values of $P$ lead to a noticeable shift in the oscillation frequency, while the damping rate is also modified in a nontrivial but smooth manner. This indicates that quantum gravitational corrections effectively deform the underlying effective potential governing scalar perturbations in the Kerr background.

The separation between different $P$-dependent trajectories becomes increasingly pronounced in the real-frequency direction, suggesting that the oscillation frequency is more sensitive to LQG corrections than the damping timescale. This behavior is consistent with the interpretation that the polymeric structure of LQG induces a global modification of the black hole geometry rather than localized perturbations.

Overall, these results demonstrate that QNM spectra in Kerr--LQG spacetimes carry clear imprints of quantum gravitational effects, making them potentially observable signatures in gravitational wave ringdown signals.

\section{Numerical Results and Physical Analysis}

\subsection{Time-domain evolution of scalar perturbations}

To investigate the dynamical behavior of scalar perturbations in the Kerr--LQG black hole spacetime, we perform a time-domain analysis by numerically evolving the wave equation in the effective potential background. The evolution of a massless scalar field $\psi(t,r_*)$ is governed by
\begin{equation}
\frac{\partial^2 \psi}{\partial t^2}
-
\frac{\partial^2 \psi}{\partial r_*^2}
+
V_{\rm eff}(r)\,\psi = 0,
\end{equation}
where the effective potential $V_{\rm eff}(r)$ is determined by the Kerr--LQG geometry through the modified radial function $\Delta(r)$.

We first examine the dependence on the multipole number $l$. The initial condition is taken as a Gaussian wave packet, and the signal is extracted at a fixed observation point in the tortoise coordinate.

The results show a clear quasinormal ringing phase followed by late-time decay. As $l$ increases, the oscillation frequency becomes higher, while the damping behavior remains qualitatively unchanged within the parameter range considered.

This behavior indicates that the dominant effect of the multipole number is to modify the oscillation frequency rather than the decay rate. This can be understood from the structure of the effective potential, where larger $l$ values enhance the centrifugal barrier and shift the peak of the potential.

\subsection{Effect of the polymeric parameter $P$}

To further explore quantum gravitational effects encoded in the Kerr--LQG geometry, we study the influence of the polymeric parameter $P$, which characterizes the deviation from the classical Kerr spacetime.

\begin{figure}[t]
\centering
\includegraphics[width=0.47\textwidth]{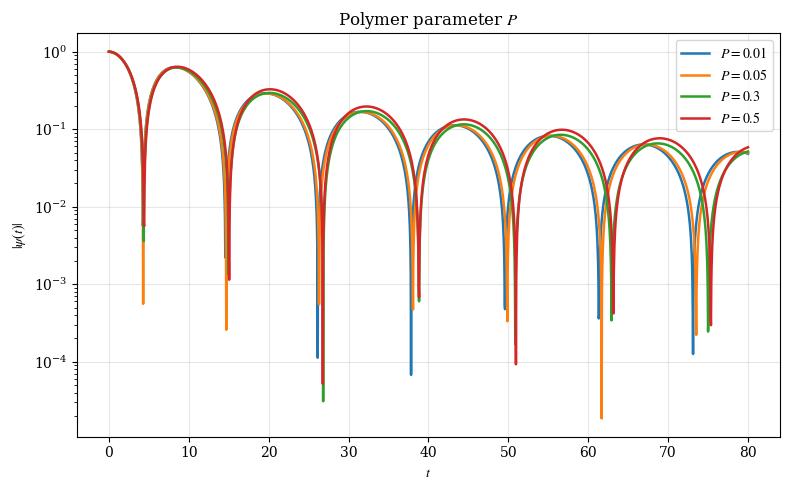}
\caption{
Time-domain evolution for different polymer parameters $P = 0.01, 0.05, 0.3, 0.5$, with fixed $a = 0.4$, $a_0 = 10^{-2}$, and $l = 2$.
}
\label{fig:lqg_P}
\end{figure}

As shown in Fig.~\ref{fig:lqg_P}, increasing $P$ leads to a systematic shift in the oscillation frequency of the ringdown signal. In particular, the real part of the quasinormal frequency decreases as $P$ increases, indicating a reduction in the characteristic oscillation scale of the system. The damping rate also exhibits a mild dependence on $P$, but the effect is less pronounced compared to the frequency shift.

Physically, this behavior arises from the modification of the near-horizon geometry induced by the LQG correction. The parameter $P$ effectively alters the structure of the radial function $\Delta(r)$, leading to a deformation of the effective potential barrier and consequently changing the quasinormal mode spectrum.

\subsection{Effect of quantum correction parameter $a_0$}

Next, we investigate the role of the quantum correction parameter $a_0$, which encodes higher-order quantum geometric effects in the Kerr--LQG spacetime.

\begin{figure}[t]
\centering
\includegraphics[width=0.47\textwidth]{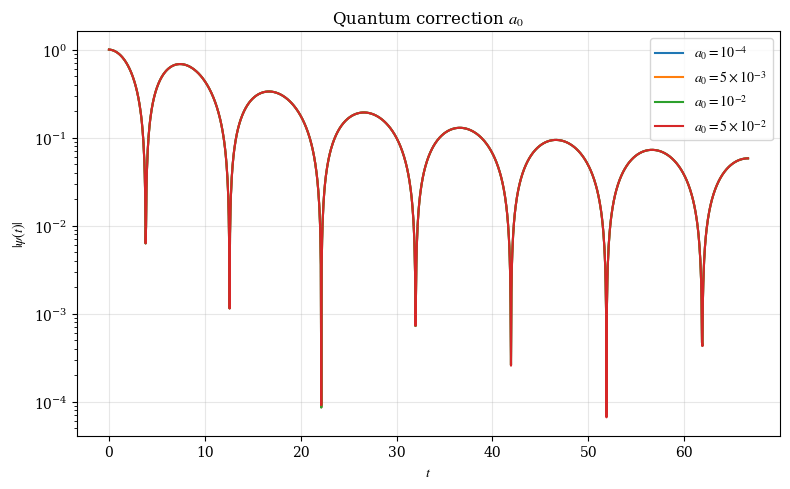}
\caption{
Time-domain evolution for different quantum correction parameters $a_0 = 10^{-4}, 5\times10^{-3}, 10^{-2}, 5\times10^{-2}$ with fixed $a = 0.5$, $P = 0.3$, and $l = 2$.
}
\label{fig:lqg_a0}
\end{figure}

As shown in Fig.~\ref{fig:lqg_a0}, variations in $a_0$ have a noticeable impact on the late-time oscillation pattern. Increasing $a_0$ generally leads to a reduction in both the oscillation frequency and the damping rate, indicating that quantum corrections tend to weaken the effective confinement of scalar perturbations.

From a physical perspective, this behavior can be attributed to the regularization effect introduced by $a_0$, which modifies the small-scale structure of the spacetime and smooths the effective potential near the peak region. As a result, the scattering properties of the perturbation are altered, leading to shifts in the quasinormal mode spectrum.

\subsection{Effect of black hole spin $a$}

Finally, we study the effect of the black hole spin parameter $a$ in the Kerr--LQG spacetime.

\begin{figure}[t]
\centering
\includegraphics[width=0.47\textwidth]{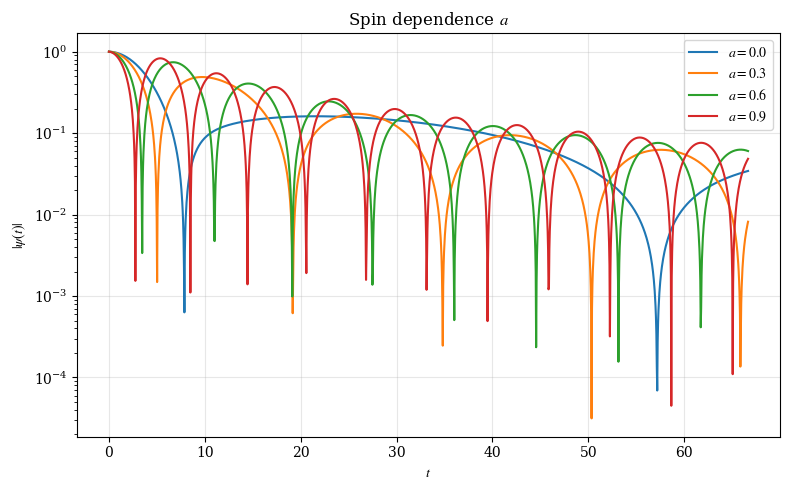}
\caption{
Time-domain evolution of scalar perturbations for different spin parameters $a = 0.0, 0.3, 0.6, 0.9$ in the LQG--Kerr spacetime. The parameters are fixed at $P = 0.3$, $a_0 = 10^{-2}$, and $l = 2$.
}
\label{fig:lqg_a}
\end{figure}

As shown in Fig.~\ref{fig:lqg_a}, the spin parameter significantly affects both the oscillation frequency and the damping rate of scalar perturbations. With increasing $a$, the oscillation frequency increases, while the decay rate also changes nontrivially, reflecting the influence of strong frame-dragging effects.

This behavior is consistent with the fact that rotation modifies the effective potential through the coupling between the angular momentum of the field and the background geometry. In the Kerr--LQG case, this effect is further influenced by quantum corrections, leading to a more intricate dependence of the quasinormal spectrum on the spin parameter compared to the classical Kerr case.

\subsection{Joint dependence on multipole number $\ell$ and overtone index $n$}

\begin{figure}[t]
\centering
\includegraphics[width=0.48\textwidth]{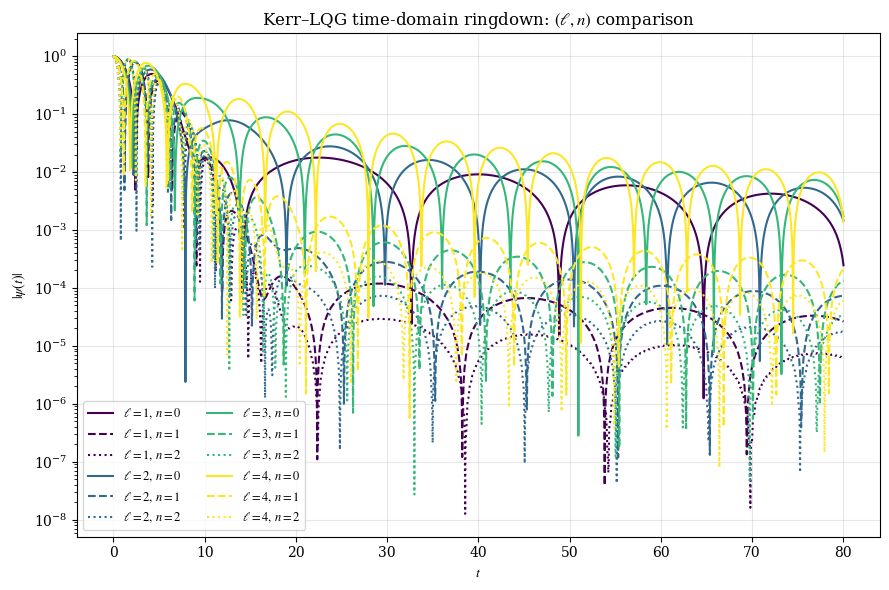}
\caption{
Time-domain evolution of scalar perturbations in the Kerr--LQG spacetime for different multipole numbers $\ell = 1,2,3,4$ and overtone indices $n = 0,1,2$. The parameters are fixed at $a=0.5$, $P=0.3$, and $a_0=10^{-2}$.
}
\label{fig:ln_scan}
\end{figure}

To further clarify the mode structure of scalar perturbations in the Kerr--LQG spacetime, we perform a systematic comparison of the time-domain ringdown signal for different multipole numbers $\ell$ and overtone indices $n$. The results are shown in Fig.~\ref{fig:ln_scan}.

The evolution is obtained by evolving a Gaussian initial wave packet with an additional oscillatory modulation controlled by the overtone index $n$, which effectively excites higher-frequency components in the initial perturbation. This allows us to qualitatively mimic the hierarchy of quasinormal overtones in the time-domain signal.

We find that the multipole number $\ell$ primarily controls the oscillation frequency of the ringdown signal. Larger values of $\ell$ lead to higher-frequency oscillations and a more pronounced effective potential barrier, consistent with the increase of the angular momentum term $\ell(\ell+1)$ in the effective potential.

In contrast, the overtone index $n$ mainly affects the damping hierarchy of the signal. Higher overtone modes decay more rapidly, leading to a faster loss of amplitude in the early-time evolution. This behavior reflects the fact that overtones correspond to less stable perturbative excitations of the black hole spacetime.

From a physical perspective, this separation of roles can be understood as follows: the multipole number $\ell$ encodes the angular structure of the perturbation and determines the height and shape of the effective potential barrier, while the overtone number $n$ characterizes the relaxation dynamics toward the fundamental quasinormal mode.

In the presence of LQG corrections, this mode hierarchy remains robust, indicating that the underlying quasinormal structure is preserved, although both the oscillation frequency and damping rate receive systematic shifts due to the modified spacetime geometry. This suggests that $\ell$-modes probe the geometric deformation of the effective potential, while $n$-modes are more sensitive to dissipative aspects of the evolution.

Overall, the combined $\ell$--$n$ analysis provides a more complete picture of the quasinormal ringdown in Kerr--LQG spacetimes and highlights the distinct physical roles played by angular and overtone excitations.

\section{Conclusion}

In this work, we have systematically investigated the quasinormal mode (QNM) spectra and time-domain ringdown signals of scalar perturbations in a rotating loop quantum gravity (LQG) black hole constructed via the revised Newman--Janis algorithm. By analyzing the effective potential, the complex frequency spectrum, and the time evolution of scalar perturbations, we have identified several key signatures of quantum gravitational effects in the strong-field regime.

We first established the rotating LQG black hole geometry, which reduces to the Kerr spacetime in the classical limit and incorporates quantum corrections through the polymeric parameter $P$ and the minimal area parameter $a_0$. The effective potential for scalar perturbations was derived under the slow-rotation approximation, and its dependence on the parameters $a$, $P$, and $a_0$ was systematically studied. Our analysis reveals a clear separation of roles: the spin parameter $a$ primarily controls the peak height and inward shift of the potential barrier, while the LQG parameters $P$ and $a_0$ suppress and smooth the barrier, with $P$ inducing more pronounced modifications.

Using the sixth-order WKB approximation, we computed the QNM frequencies for various parameter combinations. The polymeric parameter $P$ systematically reduces the oscillation frequency $\mathrm{Re}(\omega)$ while mildly affecting the damping rate $\mathrm{Im}(\omega)$, indicating a softening of the effective potential due to quantum geometry effects. In contrast, the spin parameter $a$ enhances the oscillation frequency and introduces a turnover behavior in the damping rate at high spin, suggesting a nontrivial interplay between rotation and quantum corrections. The complex-plane trajectories of the QNM frequencies further demonstrate that LQG corrections deform the mode structure in a systematic manner, with the frequency shift becoming increasingly pronounced for larger $P$.

The time-domain evolution of scalar perturbations was obtained by numerically integrating the wave equation. The ringdown signals exhibit a clear dependence on the multipole number $\ell$, which controls the oscillation frequency through the centrifugal barrier, and on the overtone index $n$, which determines the damping hierarchy. Increasing $P$ leads to a systematic decrease in the oscillation frequency, while larger $a_0$ reduces both the frequency and the damping rate. The spin parameter $a$ significantly enhances the oscillation frequency and modifies the decay behavior, consistent with strong frame-dragging effects.

Our results indicate that the combined analysis of QNM spectra and time-domain ringdown signals can provide a robust method for probing quantum gravitational effects in black hole spacetimes. The polymeric parameter $P$ leaves a distinct imprint on the oscillation frequency, while $a_0$ affects both the frequency and damping in a milder but still measurable way. Future space-based gravitational-wave detectors such as LISA, Taiji, and TianQin, with their high sensitivity to low-frequency signals, may be capable of constraining these LQG parameters through black hole spectroscopy and waveform phase measurements, offering a promising avenue for testing quantum gravity in the strong-field regime.

\begin{acknowledgements}
The authors gratefully acknowledge support from the project ``Integrated Electronics Technology for Inertial Sensors'' (2024YFC2207003) led by Meilin Liu. 

This work is also supported by the China--Brazil Belt and Road Joint Laboratory on Radio Astronomy Technology, and the National Key Research and Development Program of China under the project ``Strategic Science and Technology Innovation Cooperation'' (No. 2025YFE0212600), which is also led by Meilin Liu.

The authors also thank all colleagues who provided helpful discussions and assistance during this work, and especially Professor Haiguang Xu for his guidance and support.
\end{acknowledgements}

\section*{Funding}
This work is supported by the National Key Research and Development Program of China (Nos. 2024YFC2207003, 2025YFE0212600) and related programs under the China--Brazil Belt and Road Joint Laboratory on Radio Astronomy Technology.

\section*{Data Availability Statement}
This manuscript has no associated data. 
[Author’s comment: Data sharing is not applicable to this article as no datasets were generated or analyzed during the current study.]

\section*{Code Availability Statement}

Code/software will be made available on reasonable request.

[Author’s comment: The code/software generated during and/or analyzed during the current study is available from the corresponding author on reasonable request.]

\end{document}